\documentclass{mn2e}[3]
\usepackage{rotating}
\input psfig.sty
\begin{document}

\title[binary evolution \& blue stragglers]
{Binary coalescence from case A evolution -- mergers and blue stragglers}
\author[Chen and Han]{Xuefei Chen$^{1}$\thanks{
xuefeichen717@hotmail.com} and Zhanwen Han$^{1}$\\
$^1$National Astronomical Observatories/Yunnan Observatory, CAS, Kunming,
650011, P.R.China}
\maketitle

\begin{abstract}
We constructed some main-sequence mergers from case A binary
evolution and studied their characteristics via Eggleton's stellar
evolution code. Both total mass and orbital angular momentum are
conservative in our binary evolutions.
 Assuming that the matter from the secondary homogeneously mixes with
the envelope of the primary and that no mass are lost from the
system during the merger process, we found that some mergers might
be on the left of the zero-age main sequence as defined by normal
surface composition (i.e helium content $Y=0.28$ with metallicity
$Z=0.02$ for Pop I) on a colour-magnitude diagram(CMD) because of
enhanced surface helium content. The study also shows that central
hydrogen content of the mergers is independent of mass. Our simple
models provide a possible way to explain a few blue stragglers
(BSs) observed on the left of zero-age main sequence in some
clusters, but the concentration toward the blue side of the main
sequence with decreasing mass predicted by Sandquist et al. will
not appear in our models. The products with little central
hydrogen in our models are probably subgiants when they are
formed, since the primaries in the progenitors also have little
central hydrogen and will likely leave the main sequence during
merger process. As a consequence, we fit the formula of magnitude
$M_{\rm v}$ and $B-V$ of the mergers when they return back to
thermal equilibrium with maximum error 0.29 and 0.037,
respectively.

Employing the consequences above, we performed Monte Carlo
simulations to examine our models in an old open cluster NGC 2682
and an intermediate-age cluster NGC 2660. Angular momentum loss
(AML) of low mass binaries is very important in NGC 2682 and its
effect was estimated in a simple way. In NGC 2682, binary mergers
from our models cover the region with high luminosity and those
from AML are located in the region with low luminosity, existing a
certain width. The BSs from AML are much more than those from our
models, indicating that AML of low mass binaries makes a major
contribution to BSs in this old cluster. Our models are
corresponding for several BSs in NGC 2660. At the region with the
most opportunity on the CMD, however, no BSs have been observed at
present. {\bf Our results are well-matched to the observations if
there is $\sim 0.5M_\odot$ of mass loss in the merger process, but
a physical mechanism for this much mass loss is a problem.}

\end{abstract}


\section{Introduction}
Much evidence shows that primordial binaries make an important
contribution to blue stragglers (BSs) \cite{fer03,dpa04,map04}. At
present, a few BSs, i.e. F190, $\theta$ Car, have already been
confirmed to be in binaries by observations, and their formation
may be interpreted by mass transfer between the components of a
binary. Whereas in intermediate-age and old open and globular
clusters, the number of observed close binaries among well-studied
BSs is consistent with the hypothesis of binary coalescence. For
example, Mateo et al. \shortcite{mat90} made a comparison of the
number of close binaries with the total number of BSs in NGC 5466
and found that it is an acceptable claim that all non-eclipsing
BSs are formed as the result of mergers of the components in close
binaries, though the possibility of other mechanisms to produce
BSs cannot be ruled out due to the large uncertainties in their
analysis. Monte-Carlo simulations of binary stellar evolution
\cite{pol94} also show that binary coalescence may be an important
channel to form BSs in some clusters (e.g. with an age greater
than 40 Myr). Meanwhile, the arguments in theory show that W UMa
binaries (low-mass contact binaries) must eventually merge into a
single star \cite{web76,web85,ty87,mat90}. Observationally, the
lack of radial velocity variations for most BSs further indicates
that binary coalescence may be more important than mass transfer
for BS formation \cite{str93,pol94}. FK Comae stars are generally
considered to be direct evidence for binary coalescence
\cite{str93}. The smallest mass ratio of components among observed
W UMa systems to date is about 0.06. All of the above show that it
is important to study the remnants of close binaries. However the
merge process is complicated and the physics during the process is
still uncertain. Recently, Andronov, Pinsonneault \& Terndrup
\shortcite{apt06} studied the mergers of close primordial binaries
by employing the angular momentum loss rate inferred from the
spindown of open cluster stars. Their study shows that main
sequence mergers can account for the observed number of single BSs
in M67 and that such mergers are responsible for at least  one
third of the BSs in open clusters older than 1 Gyr. The physics of
mergers are limiting case treatments in the study of Andronov,
Pinsonneault \& Terndrup \shortcite{apt06}. Based on previous
studies of contact binaries and some assumptions, we construct a
series of merger models in this paper, to study the structure and
evolution of the models and show some comparisons with
observations.

Case A binary evolution has been well studied
by Nelson \& Eggleton \shortcite{nel01}.
They defined six major subtypes for the evolution
(AD, AR, AS, AE, AL and AN) and two rare cases (AG and AB).
Three of the subtypes (AD, AR, AS) lead the binary contact as both components
are main--sequence stars
and two cases (AE and AG) reach contact
with one or both components having left the main sequence.
As there is no description for weird objects except for
two merged main--sequence stars,
merger products (except for two main-sequences stars) are generally
assumed to have terminated their evolution \cite{pol94},
i.e. they have left the main sequence and cannot be recognized as BSs.
Here we are interested in the cases of two main-sequence stars,
i.e. cases AS, AR and AD.
If $t_{\rm dyn}$, $t_{\rm KH}$, $t_{\rm MS}$ represent the dynamic timescale,
thermal timescale and main sequence timescale of the primary
(the initial massive star,*1), respectively,
the following shows a simple definition of the three evolutionary cases:
AD--dynamic Roche lobe overflow (RLOF), $\dot{M}>M/t_{\rm dyn}$;
AR--rapid evolution to contact,  $\dot{M}>M/t_{\rm KH},
t_{\rm contact} -t_{\rm RLOF}(*1)<0.1t_{\rm MS}(*1)$;
AS--slow evolution to contact,
$t_{\rm contact} -t_{\rm RLOF}(*1)>0.1t_{\rm MS}(*1)$,
where $t_{\rm RLOF}$ and $t_{\rm contact}$ are the ages
at which RLOF begins and the binary comes into contact, respectively.
In case AD,
the core of the secondary spirals in quickly
and stays in the center of the merger.
The merger then has a chemical composition similar to that of the primary,
resembling the result of smoothed particle hydrodynamic calculations
\cite{lrs96,sill97,sill01}.
We therefore studied just the systems in cases AR and AS for this work.

\section {Assumptions}
Using the stellar evolution code devised
by Eggleton \shortcite{egg71,egg72,egg73},
which has been updated with the latest physics over the last three decades
\cite{han94,pol95,pol98},
we re-calculate the models of cases AS and AR with primary masses between
0.89 and $2M_\odot$ until the systems become contact binaries.
The structures of the primaries and the compositions of the secondaries
are stored to construct the merger remnants.

Before the system comes into contact,
the accreting matter is assumed to be deposited onto
the surface of the secondary with zero falling velocity
and distributed homogeneously all over the outer layers.
The change of chemical composition on the secondary's surface caused by the
accreting matter is
\begin{equation}
{\partial X_i / \partial t }={(\partial M /\partial t)/[(\partial M /\partial t){\rm d}t+M_{\rm s}}] \cdot (X_{i{\rm a}}-X_{i{\rm s}}),
\end {equation}
where $\partial M /\partial t$ is the mass accretion rate,
$X_{i{\rm a}}$  and $X_{i{\rm s}}$ are element abundances
of the accreting matter and of the secondary's surface for species $i$,
respectively,
and $M_{\rm s}$ is the mass of the outermost layer of the secondary.
The value of $M_{\rm s}$ will change
with the moving of the non-Lagrangian mesh
as well as the chosen model resolution,
but it is so small ($\sim 10^{-9}-10^{-12} M_{\odot}$)
in comparison with
$(\partial M /\partial t){\rm d}t$ ($\sim 10^{-3}-10^{-5} M_{\odot}$)
during RLOF that we may ignore the effect of various $M_{\rm s}$
on element abundances.
Before and after RLOF, we get $\partial X_{\rm i}/\partial t =0$
from the equation,
which is reasonable in the absence of mixing \cite{ch04}.

The merger models are constructed based on the following assumptions:
(i) contact binaries with two main-sequence components coalesce finally and
the changes of structures of individual components
during coalescence are ignored;
(ii) the matter of the secondary is homogeneously mixed with
that of the primary beyond the core-envelope transition point,
which separates the core and the envelope of the mass donor;
(iii) the system mass is conserved.

Firstly, we present a brief discussion on these assumptions.
Webbink \shortcite{web76} studied the evolutionary fate of
low-mass contact binaries, and found that a system cannot sustain
its binary character beyond the limits set by marginal contact
evolution ($\mu =M_1/(M_1+M_2)=1.0$). He stated that a contact
binary will very likely coalesce as the primary is still on the
main sequence in a real system. Up to now, it is widely believed
that case AD probably leads to common envelope, spiral-in, and
coalescence on quite a short timescale. The final consequences of
AS and AR are not very clear, but Eggleton \shortcite{egg00}
pointed out that systems undergoing AR or AS evolution may
maintain a shallow contact (perhaps intermittently) as the mass
ratio becomes more extreme, and finally coalesce. Recent study on
W UMa (Li, Zhang \& Han, 2005) also shows that these systems will
be eventually coalescence. The merged timescale, i.e. the time
from a binary contact to coalescence, is important here. If it is
too long, the structures of both components will change remarkably
and the system may have not completed coalescence within the
cluster age. There are many {\bf conflicting estimates} for the
timescale, however, from observations and theoretical models of
the merger process. {\bf Early observational estimates range from}
$10^7$--$10^8$ yr in various environments \cite{van79,eggen89}.
The following study explored the average age about $5 \times 10^8$
yr \cite{van94,dry02}. Bilir et al. \shortcite{bil05} pointed out
that the age difference between field contact binaries and
chromospherically active binaries, 1.61 Gyr, is likely an upper
limit for the contact stage by assuming an equilibrium in the
Galaxy, whereas the study of W UMa by Li, Han \& Zhang
\shortcite{lhz04} suggested a much longer timescale, about 7 Gyr.
We adopt the empirically estimated values in this paper {\bf (i.e
from $5 \times 10^7$ to $1 \times 10^9$ yrs)} and ignore the
changes of structure of individual components during merger
process. For low-mass contact binaries, the common envelope is
convective \cite{web77}, and the matter in it is thus homogeneous.
If a system mimics shallow contact during coalescence, it is
reasonable to assume that the matter of the secondary mixes with
the envelope homogeneously. Van't Veer \shortcite{van97} found
that the mass loss from the system during coalescence is at a rate
of about $2 \times 10^{-10}M_\odot{\rm yr}^{-1}$ by observations.
If we consider that the coalescence time is $5 \times 10^8$ yr in
a binary, only $0.1M_\odot$ is lost from the system as the binary
finally becomes a single star. We then roughly assume that the
mass is conservative during coalescence. However mass loss might
be an important way to carry orbital angular momentum away from
the binary in this process.

Secondly, we discuss the choice of the core-envelope transition point
which separates the core and the envelope in the primary.
Many characteristics of the merger are relevant to the choice,
e.g. the chemical composition in the envelope,
evolutionary track on Hertzsprung-Russel diagram,
and some observational characteristics.
Unfortunately,
one cannot find the core-envelope transition point in a main-sequence
star as easily as in evolved stars
because the density profile,
as well as many other thermodynamic quantities
(entropy, pressure, temperature etc.), is smooth and does
not have a deep gradient for main-sequence stars.
Chen \& Han \shortcite{ch05} studied
the influences of core-envelope transition point
on the mergers of contact binaries with two main-sequence components.
They found that
one may ignore the effects which result from different choices of
the transition point on colours and magnitudes of the merger
if it is outside the nuclear reaction region of the primary,
which is commonly considered
as the nearest boundary of the secondary reaches in cases AS and AR.
In this paper, the core-envelope transition is determined as the point
within which the core produces 99 per cent of total luminosity.
This choice is generally outside the nuclear reaction regions
and has little effect on the final results.

Finally the merger remnant is constructed as follows:
it has the total mass of the system
and a chemical composition within $M_{\rm 1c}$
similar to the core of the primary.
The chemical composition in the envelope of the merger is given by
\begin{equation}
X_i=(M_{i2}+M_{i1\rm b})/(M_2+m_{\rm b}),
\end{equation}
where $M_{i2}$ and $M_{i1\rm b}$ denote the total masses of species $i$
of the secondary and of the primary's envelope, respectively.
$m_{\rm b}$ is the envelope mass of the primary.
There might be a region in which
the helium abundance is less than that of the outer region.
The matter in this region then has a lower mean molecular weight than that
in the outer region.
This results in secular instability and thermohaline mixing \cite{kip80,ulr72}.
We include it as a diffusion
process in our code \cite{ch04}.

In the models of Nelson \& Eggleton \shortcite{nel01}, both total
mass and angular momentum are conservative. It was mentioned by
the authors, however, that these assumptions were only reasonable
for a restricted range of intermediate masses, i.e spectra from
about G0 to B1 and luminosity class III-V. Observationally, some
low mass binaries with late-type components show clear signs of
magnetic activity, which indicates that the systems evolve by way
of a scenario implying angular momentum loss (AML) by magnetic
braking \cite{mes84}. Magnetized stellar winds probably do not
carry off much mass, but they are rich in angular momentum because
of magnetic linkage to the binaries. For close binaries, rotation
is expected to synchronize with orbital period, so AML is at the
expense of the orbital angular momentum, resulting in orbital
decaying and the components approaching each other. A detached
binaries, then, may become contact and finally coalesce at or
before the cluster age \cite{ste95}. There are a number of
subjects including the treatment of AML
\cite{lhz04,ste06,mk06,dek06}. For simplicity, the conservative
assumption is also adopted in our binary evolutions. In old
clusters, however, AML of low mass binaries is very important and
{\bf we estimate its importance in another way (see section 4.2).}

\section{Evolutionary Results}
A set of binaries undergoing AS and AR evolution from Nelson \&
Eggleton \shortcite{nel01} are choosen to study the
characteristics of the merger products and their connections with
blue stragglers. Table A1 gives the initial parameters of the
binary systems, their RLOF information and the structures and
evolutionary consequences of the mergers. The first three columns
contain the initial mass of the primary $M_{\rm 1i}$, the initial
mass ratio $q_{\rm i}$ (the primary to the secondary) and the
initial orbital period $P_{\rm i}$ in logarithmic, where $P_{\rm
ZAMS}$ is given by (Nelson \& Eggleton, 2001)
\begin{equation}
P_{\rm ZAMS} \approx $$0.19M_{\rm 1i}+0.47M_{\rm 1i}^{2.33}\over 1+1.18M_{\rm 1i}^2$$.
\end{equation}
The fourth and fifth columns are the ages at which
Roche lobe overflow begins ($t_{\rm RLOF}$)
and the binary comes into contact ($t_{\rm contact}$) in our calculation.
The next three columns show some system parameters at $t_{\rm contact}$,
i.e. the mass of the primary $M_{\rm 1}$,
the mass of the secondary $M_{\rm 2}$
and the orbital period $P_{\rm contact}$.
The remaining columns present the evolutionary results of the mergers:
the lifetime on the main sequence ($t_{\rm MS}$),
the central hydrogen mass fraction of the merger as constructed
($X_{\rm Hcc}$) and after adjustment ($X_{\rm Hcm}$),
surface abundances for the elements H ($X_{\rm Hs}$), He ($X_{\rm Hes}$)
and the ratio of C/N at the surface ($(C/N)_{\rm s}$).

\begin{figure}
\centerline{\psfig{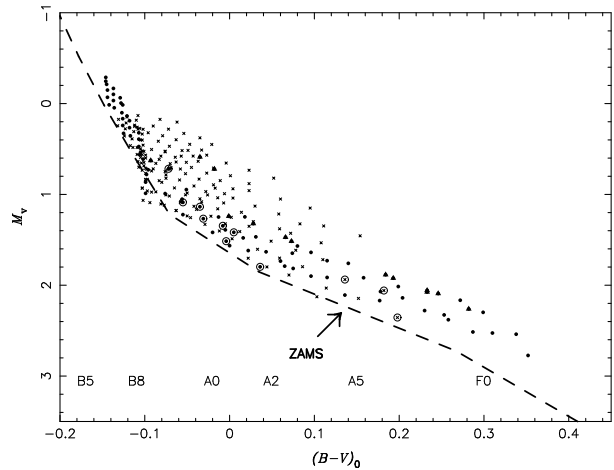}}
\caption{The locations of the mergers as they reach their
maximum hydrogen mass fraction on a colour-magnitude diagram.
`.' and `$\times $' represent the mergers from AS and AR, respectively.
Open circles show the merge models from Table 1 and the triangles are the
objects with little central hydrogen content.}
\label{CMD}
\end{figure}

Figure. \ref{CMD} shows the location of the mergers
on a colour-magnitude diagram (CMD)
when the central hydrogen mass fraction reaches its maximum,
at which we consider that
the merger returns to thermal equilibrium and begins normal evolution.
In a real case, the thermal equilibrium point of the constructed models
are probably not just at the maximum of central hydrogen mass,
but the divergence should be very small and have no influence on the results.
`.' and `$\times $' in the figure represent the mergers
from AS and AR, respectively.
Open circles show the possible BSs produced from our models
in an old open cluster M67
and the dashed one shows zero-age main sequence (ZAMS) for
$(Z,Y)=(0.02, 0.28)$,
where $Z$ is metallicity and $Y$ is helium content.
In figure \ref{CMD} we see that most of the mergers
are located in the main sequence
while a few are to the left of ZAMS.
Surface chemical composition and central hydrogen content are both
responsible for placing objects to the left of ZAMS.
Generally the mergers have a larger helium content than 0.28
because the matter from the secondary, including some helium-rich matter
produced via nuclear reaction, homogeneously mixes
in the envelope of the primary.
They are thus bluer than stars of the same mass
and of the same central hydrogen mass fraction with $Y=0.28$.
Furthermore, the mergers may be on the left of ZAMS
if central H contents close to 0.7, initial H mass fraction for Pop I.

\begin{figure}
\centerline{\psfig{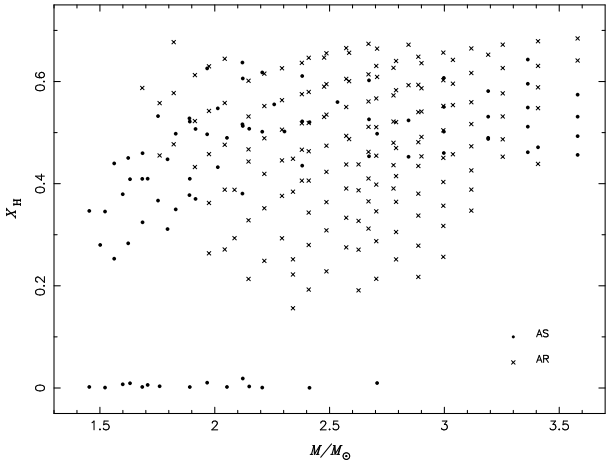}}
\caption{The maximum of central hydrogen mass fraction
for all the merger models.}
\label{XHD}
\end{figure}

Sandquist, Bolte \& Hernquist \shortcite{sbh97} argued that
a fainter BS should have a less massive progenitor,
and therefore has a lower helium content in the core.
This means that
BSs should spend an increasing amount of time
near the ZAMS with decreasing mass,
which will lead to a definite concentration
toward the blue side of the main sequence in the region of low mass BSs.
In Fig. \ref{XHD}
we present the distribution of central hydrogen mass fraction at its
maximum $X_{\rm Hcm}$ with the mass of the merger.
No evidence in the figure shows that
lower mass mergers should have higher hydrogen mass fractions
in the center (say, more close to ZAMS).
In our models, there are at least two factors affecting
the central hydrogen mass fraction of a merger, i.e.
the evolutionary degree of the parent stars
when the system comes into contact and
the development of the central convective region of the merger
before normal evolution.
The former is relevant to the mass of the progenitor
and their contact ages.
If we simply assume that all BSs in a cluster are formed at a same time
(which is obviously not true),
larger (more massive) progenitors have less hydrogen content in the center,
but their mergers may develop larger convective regions in the centers,
involving more of the H-rich matter
from around the nuclear region of the primary in the center of the merger.
As a consequence, the enhancement of central hydrogen content
might be comparable to, or larger than, the less massive mergers.
On the other hand, if a fainter BS is produced earlier than a brighter one --
the formation time of a BS is strongly dependent on
some other system's initial parameters as well as its progenitors' mass.
The fainter BS has then evolved for a relatively longer time and may
have less hydrogen in the center than the larger one.
Therefore the blue concentration with mass decreasing predicted by
Sandquist, Bolte \& Hernquist \shortcite{sbh97} may be not true in
a real case.
In fact, we have not found the concentration observationally,
whereas there is a certain width of BSs sequence.

We obtained some models with very little central hydrogen content,
i.e $X_{\rm Hcm}$ very close to zero. Seen in Fig. \ref{XHD}, all
these objects are from the AS channel. Long-time RLOFs of their
progenitors lead the primaries be near the termination of main
sequence when the systems being contact. Detailed calculations for
these mergers show that their main-sequence timescales are in
order of $10^8$ yrs. Such unusually long lifetimes are relevant to
the development of central convective core. {\bf Comparison to the
primaries of the progenitors, the products have higher masses and
will develop larger convective core masses in the following
evolutions, leading central hydrogen increase. On the other hand,
nuclear reactions will consume some hydrogen in the core. The
maximum of central hydrogen mass fraction is the equibibrium point
at which the hydrogen involved in the core is comparable to that
exhausted by nuclear reactions. After that, the consumed hydrogen
is more than that involved in the core, but the growth of
convective core adds fresh fuels and extends the lifetime in this
phase.} For a $1.6M_\odot$ star with a stable convective core, it
is just about $2\times10^7$ yr when hydrogen mass fraction ranges
from 0.005 to 0.0000, while about $10^8$ yr in these models.

We should pay more attention to the mergers with little central
hydrogen content. The triangles in figure 1 show the positions of
these objects. From Table A1, we see that the primaries are very
close to the exhausted of hydrogen. So it is very likely that the
primaries have left the main sequence as eventually coalescence,
and the mergers are subgiants and much redder than that shown in
Fig. 1. Meanwhile, the development of convective core and the
merger process are probably synchronous. After coalescence, the
convective core develops more quickly than that in our models.
Sandquist, Bolte \& Hernquist \shortcite{sbh97} once suggested
that low-mass BSs (near the turnoff of a cluster) in M67 might be
subgiant stars and a very small central hydrogen content in these
objects may remarkably shorten their MS lifetimes and increase the
chance of them being in the subgiant region. This may explain the
observed spread in colour of low-mass BSs. These mergers with
little central hydrogen in our models are similar to the objects
mentioned by Sandquist, Bolte \& Hernquist \shortcite{sbh97}, but
they may be up to $2.7M_\odot$, not just staying in the region
around the turnoff in M67.

\begin{figure}
\centerline{\psfig{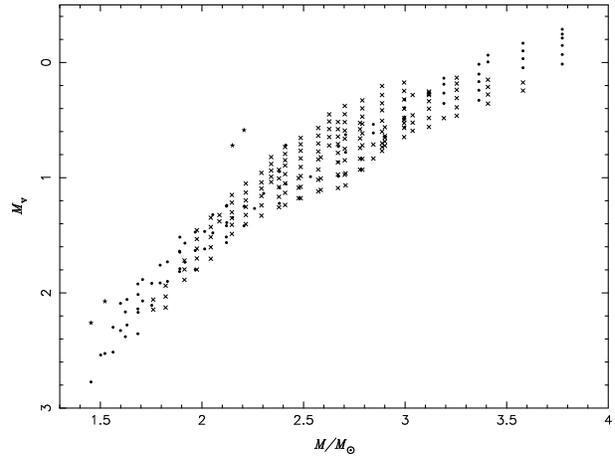}}
\caption{Visual magnitude changes with the mass of the merger.
The dots and open crosses are for AS and AR, respectively.
The stars represent the ones which are ruled out in eqs.(7) and (8).}
\label{mv}
\end{figure}

\begin{figure}
\centerline{\psfig{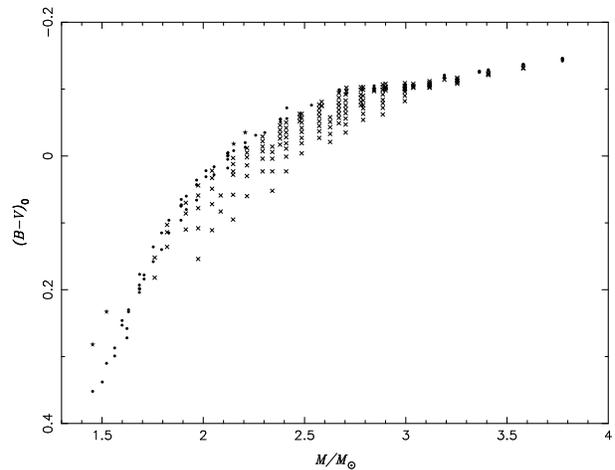}}
\caption{Colour index $B-V$ changes with the mass of the merger.
The dots and open crosses are for AS and AR, respectively.}
\label{bv}
\end{figure}

Figures.\ref{mv} and \ref{bv} show how
the magnitude in the V band, $M_{\rm V}$,
and the colour, $B-V$, change with the merger mass $M_{\rm BS}$
at $X_{\rm Hcm}$.
$M_{\rm V}$ is calculated by
\begin{equation}
M_{\rm V} = M_{\rm bol}-BC, M_{\rm bol}=4.75-2.5\times {\rm log}(L/L_{\odot}),
\end{equation}
where $BC$ and $B-V$ are obtained by linear interpolation
from the BaSel-2.0 model \cite{lcb97,lcb98}.

\begin{figure}
\centerline{\psfig{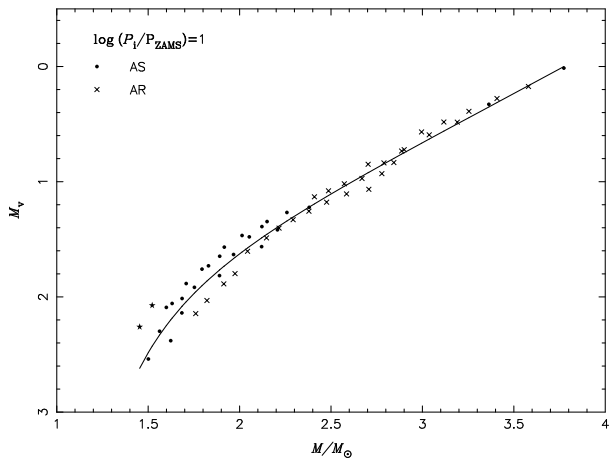}}
\caption{Visual magnitude changes with the mass of the merger as
${\rm log}(P_{\rm i}/P_{\rm ZAMS})=0.10$ .
The dots and open crosses are for AS and AR, respectively.
The stars represent the ones which are ruled out in fitting eqs.(5).
The solid line shows the fitting curve via equation(5).}
\label{mv2}
\end{figure}

\begin{figure}
\centerline{\psfig{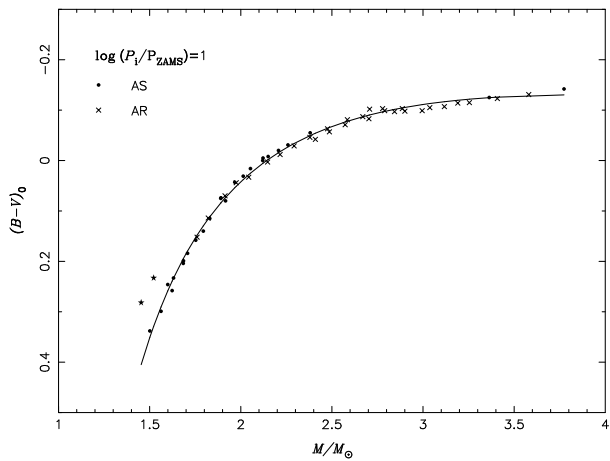}}
\caption{Colour index $B-V$ changes with the mass of the merger
as ${\rm log}(P_{\rm i}/P_{\rm ZAMS})=0.10$.
The dots and open crosses are for AS and AR, respectively.
The stars represent the ones which are ruled out in fitting eqs.(6).
The solid line shows the fitting curve via equation(6).}
\label{b-v2}
\end{figure}

As seen in the figures,
$M_{\rm V}$ and $B-V$ are affected
not only by the total mass, but also by the initial orbital period.
For ${\rm log}P_{\rm i}/P_{\rm ZAMS}=0.1$,
Figs. \ref{mv2} and \ref{b-v2} present $M_{\rm V}$ vs $M_{\rm BS}$ and
 $B-V$ vs $M_{\rm BS}$, respectively.
The solid lines are fitted by
(two models marked with star are ruled out
because they are outliers):

\begin{equation}
M_{\rm v}=$$-10.93+22.33M-2.65M^{2.5}\over -10+10.76M$$
\end{equation}

\begin{equation}
(B-V)_0=$$47.89+9.75M-46.99M^{0.5}\over -10+16.07M$$
\end{equation}
The maximum errors for the two equations are 0.19 and 0.019,
respectively.

\begin{figure}
\centerline{\psfig{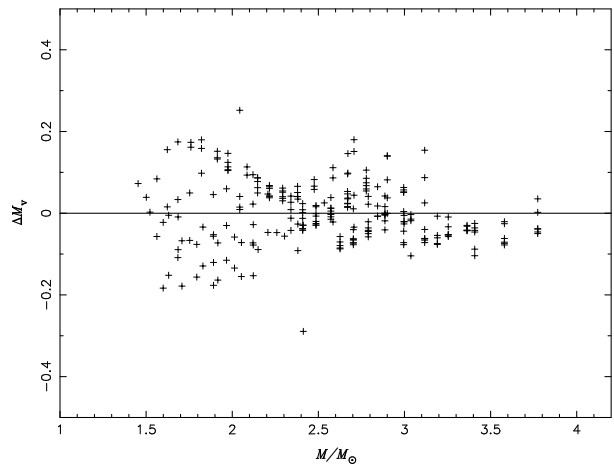}}
\caption{Visual magnitude error from equation (6) and the calculations.}
\label{mverr}
\end{figure}

\begin{figure}
\centerline{\psfig{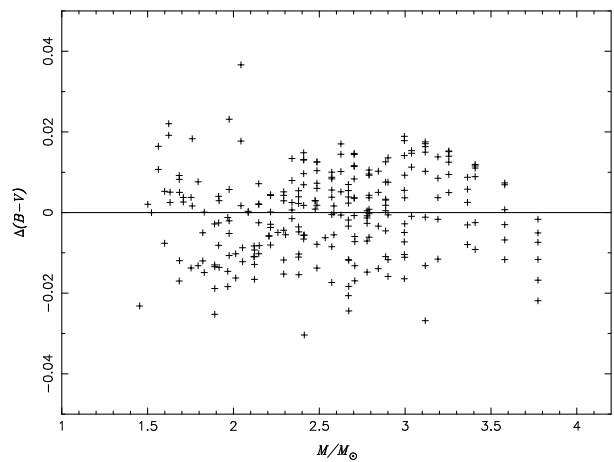}}
\caption{Colour index $B-V$ error  from equation (7) and the calculations.}
\label{bverr}
\end{figure}

Considering the initial orbital period, we get
\begin{equation}
M_{\rm v}=$$15.87+19.58M-3.01M^{2.5}\over -10+16.22M$$-1.5({\rm log}P_{\rm i}/P_{\rm ZAMS}$$-0.05)
\end{equation}

\begin{equation}
(B-V)_0=$$102+21.73M-101.5M^{0.5}\over -10+27.44M$$+$$623({\rm log}P_{\rm i}/P_{\rm ZAMS}-0.05)^{2} \over 10+62.4M^{3}$$
\end{equation}
The maximum errors for the two equations are 0.29 and 0.037, respectively.
The four models marked with stars in Figs.\ref{mv} and \ref{bv}
are also ruled out in the fitting for the same reason as above.
The distribution of error for eqs.(6) and (7), from detailed calculation,
are shown in Figs.\ref{mverr} and \ref{bverr}.

Initial parameters of the four systems ruled out in eqs.(6) and (7)
are ($M_{\rm 1i}, {\rm log}q_{\rm i},{\rm log}(P_{\rm 0}/P_{\rm ZAMS})$)=
(0.89, 0.2, 0.1), (0.89, 0.25, 0.1), (1.26, 0.15, 0.15) and (1.41, 0.25, 0..2).
From Table A1, we see that all four of these mergers
have central hydrogen content close to zero,
i.e. 0.0007, 0.0020, 0.0028 and 0.0007, respectively.
This might be the reason that they deviate from most of our models

\begin{figure}
\centerline{\psfig{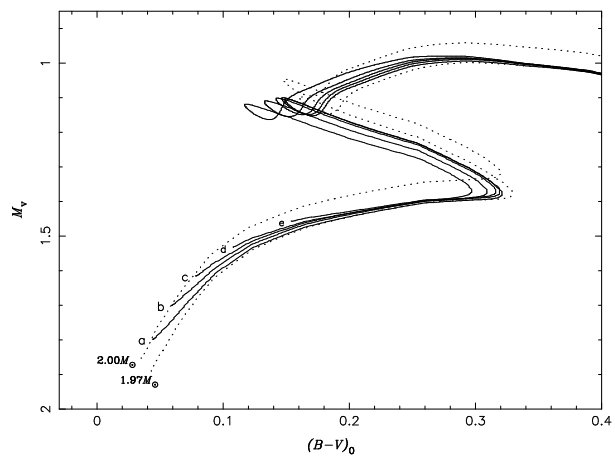}}
\caption{Evolutionary tracks of the mergers after adjustment.
($M_{\rm 1i}, {\rm log}q_{\rm i}$)=(1.41, 0.40) for the solid lines and
a, b, c, d, e represent
${\rm log}P_{\rm i}/P_{\rm ZAMS}=0.10,0.15, 0.20, 0.25,0.30$, respectively.
The dotted lines show the evolution of two stars with
surface helium content $Y$=0.28 for $M_{\rm 1i} =1.97$ and $2.00M_{\odot}$.}
\label{197CMD}
\end{figure}

The evolutionary tracks of the mergers with $M=1.97M_{\odot}$
(($M_{\rm 1i}, {\rm log}q_{\rm i}$)=(1.41, 0.40))
are shown in Fig. \ref{197CMD}.
Two stars with $Y=0.28$ and $M=2.00$ and $1.97M_{\odot}$ are also presented
in the figure (dotted lines) for convenient comparison.
In the figure,
we can hardly distinguish the difference from various initial orbital period
except for the main-sequence life, $t_{\rm MS}$, of the mergers.
Though $t_{\rm MS}$ cannot be directly observed,
it may affect observations, i.e.
together with $t_{\rm contact}$ of the progenitor system,
it is crucial for whether or not the merger can be observed or not by now.
At the same time, various initial orbital periods result in
different luminosity functions for mergers --
a longer initial orbital period leads a shorter $t_{\rm MS}$,
but during the shorter $t_{\rm MS}$,
the merger mainly stays in the relatively high luminosity region
(Fig. \ref{197CMD}).

\begin{figure}
\centerline{\psfig{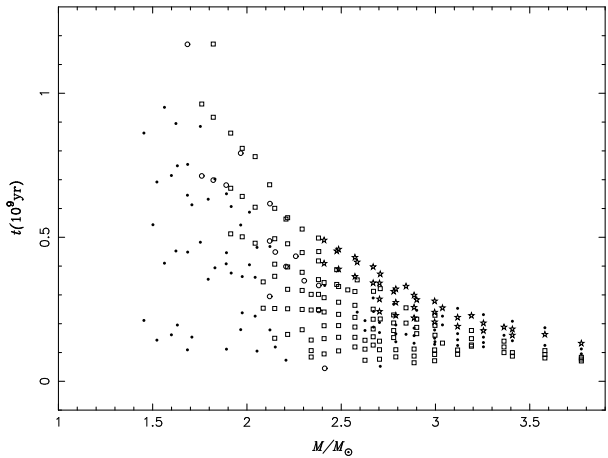}}
\caption{The timescale of the mergers staying on the main sequences.
As $t_{\rm cc}=5 \times 10^8$ yr, the open circles show the models fulfilled
$t_{\rm contact}+t_{\rm cc}\le 4.5\times 10^9$ yr and
$t_{\rm contact}+t_{\rm cc}+t_{\rm MS}\ge 3.5\times 10^9$ yr.
The squares and stars are for
$t_{\rm contact}+t_{\rm cc}\le 2.4\times 10^9$ yr and
$t_{\rm contact}+t_{\rm cc}+t_{\rm MS}\ge 1.4\times 10^9$ yr and
$t_{\rm contact}+t_{\rm cc}\le 1.0\times 10^9$ yr and
$t_{\rm contact}+t_{\rm cc}+t_{\rm MS}\ge 0.9\times 10^9$ yr, respectively.}
\label{tms}
\end{figure}

\section{the mergers and BSs in open clusters}
If a BS in a cluster is the merger remnant of binary coalescence
\footnote{It just means AS and AR evolutionary channels in this section},
then there are some constrains on the progenitor and on the merger itself.
Firstly the total mass of the progenitor should be larger than
the turnoff of the cluster $M_{\rm to}$.
Secondly, the contact age of the progenitor $t_{\rm contact}$
should be appropriate.
It cannot be so short that the merger has terminated its
evolution on the main sequence, indicating
$t_{\rm contact}+t_{\rm cc}+t_{\rm MS}\ge t_{\rm cluster}$.
Here $t_{\rm cc}$ is the timescale from binary contact to final merger.
$t_{\rm contact}$ also cannot be too long
to complete its merger process at $t_{\rm contact}$, indicating
$t_{\rm contact}+t_{\rm cc}\le t_{\rm cluster}$.
If a binary has already come into contact
but has not completed the mergering process
(i.e. a W UMa system), it can also be considered as a BS \cite{str93},
e.g. S1036 and S1282 in M67.
The initial parameter space for W UMa systems may be obtained by
$t_{\rm contact}+t_{\rm cc}\ge t_{\rm cluster} \ge t_{\rm contact}$.

\begin{table}
 \begin{minipage}{85mm}
 \caption{Characteristics of some open clusters (i.e. the age $t$
and metallicity $Z$)
 and the numbers of BSs ($N_{\rm BS}$), W UMa
systems ($N_{\rm W UMa}$) and stars on the main sequence to two
magnitudes below the turnoff ($N_{\rm 2}$) in them. Most of the
information is from Rucinski(1998) for $N_{\rm W UMa}$, Ahumada \&
Lapasset (2007) for $N_{\rm BS}$ and $N_{\rm 2}$, and Xin \&
Deng(2005) for $t$ and $Z$. The data not from the references above
are marked with 1, 2, where 1 -- Mochejska et al. (2004); 2 --
Kafka et al.(2004). The stars means that we have not found related
reports.}
 \label{tab1}
   \begin{tabular}{lccccc}
\hline
ID & $t (Gyr)$ & $Z$ & $N_{\rm bs}$ & $N_{\rm W UMa}$ & $N_{\rm 2}$\\
\hline
Be 33    & 0.7 & 0.005 & 2  & 1 & 270\\
Tom 2    & 1   & 0.009 & 17 & 4 & 440\\
NGC 2243 & 1.1 & 0.007 & 9  & 2 & 120\\
NGC 2158 & 1.2 & 0.006 & 40 & $6^1$ & 700\\
NGC 2660 & 1.2 & 0.02  & 8  & $0^*$ & 150\\
NGC 6939 & 1.6 & 0.02  & 5  & $6^2$ & 180\\
NGC 3680 & 1.6 & 0.026 & 1  & $0^*$ & 30\\
NGC 752  & 1.7 & 0.014 & 1  & 1 & 25\\
NGC 7789 & 2.0 & 0.016 & 22 & 5 & 130\\
NGC 2682 & 4.0 & 0.02  & 30 & 3 & 175\\
NGC 188  & 7.0 & 0.024 & 24 & 7 & 185\\
NGC 6791 & 7.2 & 0.039 & 75 & 4 & 800\\
Be 39    & 8.0 & 0.01  & 43 & 9 & 600\\
\hline
\label{oc}
\end{tabular}
\end{minipage}
\end{table}

Figure \ref{tms} presents the lives of the mergers on the main
sequence. We see that some low-mass BSs (i. e. $M \le
2.0M_{\odot}$) may exist for about $10^9$ yr, which is long enough
to be observed. In most cases, it is in order of $10^8$ yr, which
is similar to that of W UMa stars from observations, and therefore
we may roughly estimate the contribution to BSs from AS and AR via
the number of W UMa systems in a cluster. The estimation, however,
is not absolutely since both of the two timescales have wide
ranges and large uncertainties, and we cannot rule out {\bf other
methods for creating W UMa systems} except for AS and AR. Table 1
presents characteristics of some open clusters and the numbers of
BSs and W UMa systems in them from observations. We see that
$N_{\rm BS}>>N_{\rm W UMa}$ in old clusters (i.e $t \ge 2.0$Gyr)
and $N_{\rm BS} \sim N_{\rm W UMa}$ in half of the left clusters,
indicating that our models (binary coalescence from AS and AR) are
not important  for the produce of BSs in old open clusters, while
likely play a critical role in some younger open clusters. In old
open clusters, where stellar collisions may be ignored because of
low stellar density, AML of low mass binaries is possibly
dominated in producing BSs, since the time is long enough for
binaries with initial orbital period about 2 d evolving from
detached to contact by AML and the mergers may be more massive
than the turnoff. Meanwhile, from initial mass functions which
have been presented, most stars are concentrated on low mass.
Since the individual components almost have not evolved before
contact, their mergers from this way have much longer timescales
on the main sequences.

\subsection{Binary Samples}
To investigate BSs resulting from binary coalescence,
we have performed a Monte Carlo simulation
where a sample of $10^6$ binaries are considered
(very wide binaries are actually single stars)
including BSs originated from AS and AR evolution channels.
A single starburst is assumed in the simulation,
i.e. all the stars have the same age and metallicity ($Z=0.02$).
The initial mass function (IMF) of the primary,
the initial mass ratio distribution
and the distribution of initial orbital separation are as follows:

i) the IMF of Miller \& Scalo \shortcite{ms79} is used
and the primary mass is generated from the formula of
Eggleton, Fitchett \& Tout \shortcite{egg89}:
\begin{equation}
M_{\rm 1}=$$0.19X\over (1-X)^{0.75}+0.032(1-X)^{1/4}$$
\end{equation}
where $X$ is a random number uniformly distributed between 0 and
1. The mass ranges from 0.1 to 100$M_\odot$.

ii)the mass ratio distribution is quite controversial and,
for simplity, we only consider a
constant mass ratio distribution \cite{maz92}.
\begin{equation}
n(q)=1,  0\le q \le 1
\end{equation}
where $q=M_2/M_1$.

iii)We assume that all stars are members of binary systems
and the distribution of separations  is constant in log$a$
($a$ is separation).
\begin{equation}
an(a)=\left\{
\begin{array}{ll}
\alpha_{\rm sep}(a/a_0)^m, &a \le a_0\\
\alpha_{\rm sep}, & a_0<a<a_1\\
\end{array}
   \right.
\end{equation}
where $\alpha =0.070, a_0=10R_{\odot},a_1=5.75\times 10^6R_{\odot}=0.13pc$ and
$m=1.2$. This distribution gives an equal number of wide binary systems per
logarithmic interval and 50 per cent of systems with orbital periods of
less than 100 yr.

\subsection{NGC 2682}
Some studies show that the metallicity of NGC 2682 is a little different
from the solar \cite{car96,fan96},
while some other studies concur that it is virtually indistinguishable from
solar \cite{hob91,fri93}.
So we select this cluster as the first sample to examine our models.
The distance modulus $m-M=9.55$
and the reddening $E(B-V)=0.022$ \cite{car96,fan96}
as we translate the theory results to observations.

Previous studies on the age of this cluster showed several discrepancies.
It may range from 3.2{\underline +}0.4 Gyr \cite{bb03} to 6.0 Gyr \cite{jp94}.
The study of VandenBerg \& Stetson \shortcite{vs04} derived an age of 4.0 Gyr.
In the N-body model of this cluster \cite{hur05},
the authors investigated the behaviour around 4 Gyr.
As mentioned in section 2,
it is also a great uncertainty for the timescale $t_{\rm cc}$.
Table 2 and Table 3 present initial parameter spaces
from different considerations for both BSs and W UMa systems in our grid.
In Table 2,
we fixed the cluster age at $t_{\rm cluster}=3.8 \times 10^9$ yr
(corresponding to the turnoff of $1.26M_\odot$)
and varied $t_{\rm cc}$, i.e  $t_{\rm cc}=1 \times 10^8$ yr,
$t_{\rm cc}=5 \times 10^8$ yr and $t_{\rm cc}=1 \times 10^9$ yr.
In Table 3, we set $t_{\rm cc}=5 \times 10^8$ yr
while the age of the cluster has a width from 3.2 to 4.2 Gyr.
then the conditions for valid parameter space to form BSs via case AS or AR
are
$t_{\rm contact}+t_{\rm cc}\le 4.2\times 10^9$ yr,
$t_{\rm contact}+t_{\rm cc}+t_{\rm MS}\ge 3.2\times 10^9$ yr,
and the constrains for W UMa systems in the cluster are
$t_{\rm contact}\le 4.2\times 10^9$ yr and
$t_{\rm contact}+t_{\rm cc}\ge 3.2\times 10^9$ yr.

In Table \ref{1},
we see that $t_{\rm cc}$ will remarkably affect on initial parameter space
for both of the two kinds of objects.
For example, the initial parameter space is much larger for
$t_{\rm cc}=1 \times 10^9$ yr
than those with other $t_{\rm cc}$.
We should bear in mind that, however,
the long $t_{\rm cc}$ may be unreasonable from both of observations
and theories (see section 2).
The long $t_{\rm cc}$ may delay the appearance of the mergers
and shorten their timescales on the main sequence,
since individual evolution during mergering process are ignored.
Especially for the primaries with a very littler hydrogen  in the center
at the system contact,
the mergers may have never been on the main sequence.

From Table \ref{2}, we see that,
because of the existence of an age range,
some models might be either mergers or W UMa systems.
For large uncertainties of the age of the cluster,
the initial parameter defined this way
include almost all the models in Table \ref{1}.

\begin{table}
 \begin{minipage}{70mm}
 \caption{Initial parameters for the mergers and W UMa systems in the grid for the old open cluster M67. We take $t_{\rm cluster}=3.84\times 10^9 {\rm yr}$, corresponding turn-off is $1.26M_\odot$.}
 \label{tab2}
   \begin{tabular}{cccc}
\hline
&$t_{\rm cc}=1.0\times 10^8$ yr&\\
\hline
 &$\scriptstyle M_{\rm 1i}$ & $\scriptstyle {\rm log}q_{\rm i}$ &
   $\scriptstyle {\rm log}P_{\rm i}$\\
\hline
the mergers &  1.26 & 0.15 & 0.1\\
&&&\\
W UMa       &  1.26 & 0.05 & 0.2\\
\hline
&$t_{\rm cc}=5.0\times 10^8$ yr&&\\
\hline
the mergers &  1.26&  0.25&  0.05\\
            &  1.41&  0.15&  0.2\\
&&&\\
W UMa       &  1.26&  0.05&  0.2\\
            &  1.26&  0.15&  0.1\\
\hline
&$t_{\rm cc}=1.0\times 10^9$ yr&&\\
\hline
the mergers&1.12&  0.05&  0.05\\
           &1.12&  0.3 &  0.05\\
           &1.26&  0.05&  0.15\\
           &1.26&  0.1 &  0.1\\
           &1.26&  0.25&  0.05\\
           &1.26&  0.35&  0.15\\
           &1.26&  0.4 & 0.15\\
           &1.41&  0.2 & 0.15\\
           &1.41&  0.3 & 0.15\\
&&&\\
W UMa      &1.26&  0.05&  0.2\\
           &1.26&  0.15&  0.1\\
           &1.41&  0.15&  0.2\\
           &1.41&  0.25&  0.15\\
\hline
\label{1}
\end{tabular}
\end{minipage}
\end{table}

\begin{table}
 \begin{minipage}{70mm}
 \caption{Initial parameters for the mergers and W UMa systems in the grid
for the old open cluster M67
($t=3.2-4.2$ Gyr and $t_{\rm cc}=5\times 10^8$ yr). }
 \label{tab2}
   \begin{tabular}{ccccccccc}
\hline
& the mergers &&&&& W UMa&&\\
\hline
 $\scriptstyle M_{\rm 1i}$ & $\scriptstyle {\rm log}q_{\rm i}$ &
   $\scriptstyle {\rm log}P_{\rm i}$&&&
$\scriptstyle M_{\rm 1i}$ & $\scriptstyle {\rm log}q_{\rm i}$ &
   $\scriptstyle {\rm log}P_{\rm i}$\\
\hline
  1.12&  0.05&  0.05&&&   1.12&  0.1&  0.05\\
  1.12&  0.3&  0.05&&&  \\
  1.26&  0.05&  0.15&&&  1.26&  0.05&  0.15\\
  1.26&  0.1&  0.1&&&  1.26&  0.05&  0.2\\
  1.26&  0.15& 0.1&&& 1.26&  0.1&  0.1\\
  1.26&  0.25&  0.05&&& 1.26&  0.15&  0.1\\
  1.26&  0.35&  0.15&&& 1.26&  0.3&  0.1\\
  1.26&  0.4&  0.15&&&  \\
  1.41&  0.05&  0.25&&& 1.41&  0.15&  0.2\\
  1.41&  0.15&  0.2&&&  1.41&  0.2&  0.15\\
  1.41&  0.2&  0.15&&&  1.41&  0.25&  0.15\\
  1.41&  0.25&  0.15&&& 1.41&  0.25&  0.2\\
  1.41&  0.3&  0.15&&&   1.41&  0.3&  0.2\\
  1.41&  0.35&  0.3&&& \\
  1.41&  0.4&  0.3&&& \\

\hline
\label{2}
\end{tabular}
\end{minipage}
\end{table}

Some models with $M_{\rm 1i}=1.41M_\odot$ appear in Table \ref{1}.
We checked the evolutionary details of them in Table A1
and find that
the system
($M_{\rm 1i}, {\rm log}q_{\rm i},{\rm log}(P_{\rm 0}/P_{\rm ZAMS})$)=
(1.41, 0.15, 0.2) begins RLOF at  1.4 Gyr and becomes contact at 3.3 Gyr.
As $t_{\rm cc}=5\times10^8$ yrs, the merger is formed at 3.8 Gyr with
$X_{\rm Hcm}=0.0003$ after self-adjustment
and will leave the main sequence after 0.45 Gyr.
As discussed in section 3,
the merger of this system is very likely to be a subgiant when it formed.
For other binaries with $M_{\rm 1i}=1.41M_\odot$ in Table \ref{1},
though the primary is not near to zero,
relatively long contact timescales ($1\times10^9$)
lead the mergers be formed at about 3.5 to 3.8 Gyr.
So they are still on the main sequence at the cluster age.

Now we consider another case
which is related to the BSs in a binary BS+MS in this cluster.
It means that the secondary is a BS
before the system being contact.
The parent binary for this case should fulfill the conditions as follows:
(a) $t_{\rm contact} > t_{\rm cluster}> t_{\rm RLOF}$;
(b) at $t_{\rm cluster}$, $M_{\rm 2}> M_{\rm to}$.
Only one model is found in our results to possibly produce BS+MS in M67.
The initial parameters for the model are
$(M_{\rm 1i},{\rm log}q_{\rm i}, {\rm log}P_{\rm i})=(1.41,0.25,0.2)$.
At the age $3.82 \times 10^9$ yr, the secondary is $1.26M_\odot$
and the system comes into contact at $4\times 10^9$ yr
with an orbital period of 0.5276 d.
It means that the BS+MS may exist at least for $2 \times 10^8$yr.
A system like this should be near the turn-off if only the secondary (BS)
is considered. However no observed BSs in this region are located in a binary
like this. The evolved component may contribute to some luminosity
\cite{tian06}.

\begin{table}
 \begin{minipage}{70mm}
 \caption{Consequences of Monte Carlo simulation at different conditions
for binaries undergoing AS and AR evolutionary channels in NGC 2682 (M67).}
   \begin{tabular}{cccc}
\hline
& conditions & $N_{\rm BS}$&$N_{\rm W UMa}$\\
\hline
case 1 & $t_{\rm cc}=1 \times 10^8$yr & 5 &3 \\
       & $t_{\rm cluster}=4\times 10^9$ yr&&\\
\hline
case 2 & $t_{\rm cc}=5 \times 10^8$yr & 5 & 10\\
       & $t_{\rm cluster}=4 \times 10^9$ yr&&\\
\hline
case 3 & $t_{\rm cc}=1 \times 10^9$yr & 14& 15\\
       & $t_{\rm cluster}=4 \times 10^9$ yr&&\\
\hline
case 4 & $t_{\rm cc}=1 \times 10^8$yr & 20&13 \\
       & $t_{\rm cluster}=3.4-4.2\times 10^9$ yr&&\\
\hline
case 5 & $t_{\rm cc}=5 \times 10^8$yr & 25&19\\
       & $t_{\rm cluster}=3.4-4.2 \times 10^9$ yr&&\\
\hline
case 6 & $t_{\rm cc}=1 \times 10^9$yr & 38 & 36 \\
       & $t_{\rm cluster}=3.4-4.2 \times 10^9$ yr&&\\
\hline
\label{3}
\end{tabular}
\end{minipage}
\end{table}

\begin{table}
 \begin{minipage}{70mm}
 \caption{AML Consequences for NGC 2682.}
   \begin{tabular}{cccc}
\hline
& conditions & $N_{\rm BS}$&$N_{\rm W UMa}$\\
\hline
case 1 & $t_{\rm cc}=1 \times 10^8$yr & 855 &14 \\
       & $t_{\rm cluster}=4\times 10^9$ yr&&\\
\hline
case 2 & $t_{\rm cc}=5 \times 10^8$yr & 964 & 84\\
       & $t_{\rm cluster}=4 \times 10^9$ yr&&\\
\hline
case 3 & $t_{\rm cc}=1 \times 10^9$yr & 1079& 204\\
       & $t_{\rm cluster}=4 \times 10^9$ yr&&\\
\hline
case 4 & $t_{\rm cc}=0 \times 10^8$yr & 824 &0 \\
       & $t_{\rm cluster}=4 \times 10^9$ yr&&\\
\hline
case 5 & $t_{\rm cc}=5 \times 10^8$yr & 996&120\\
       & $t_{\rm cluster}=3.5 \times 10^9$ yr&&\\
\hline
case 6 & $t_{\rm cc}=5 \times 10^8$yr & 1194 & 109 \\
       & $t_{\rm cluster}=4.5\times 10^9$ yr&&\\
\hline
\label{aml}
\end{tabular}
\end{minipage}
\end{table}

\begin{figure*}
\centerline{\psfig{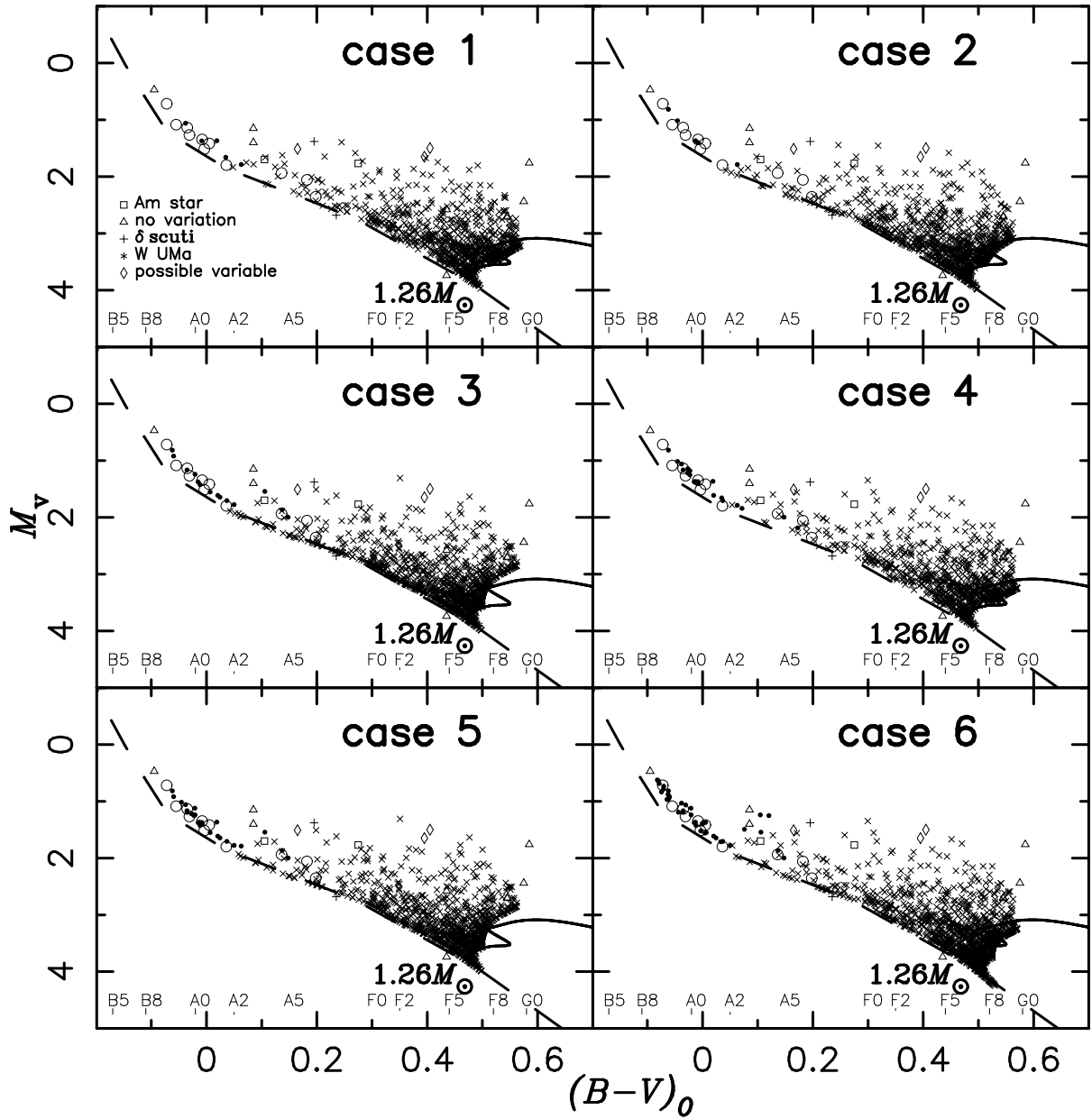}}
\caption{Monte-Carlo simulation results for M67 at different conditions
(Table \ref{3}.) `$\times $' represents the models from grid calculation,
open circles show the models in Table \ref{1} and the dots give the
possible BSs. The observed BSs from  Sandquist \& Shetrone (2003)
are also plotted in the figure.}
\label{montec}
\end{figure*}

By interpolating from Table A1,
we get different numbers of BS ($N_{\rm BS}$) under various conditions
(Table \ref{3}).
The location of the produced BSs on a CMD are presented in Fig.\ref{montec}.

As mentioned before, AML of low mass binaries probably makes a
major contribution to BSs in this old open cluster. For low mass
binaries, the individual components have almost not evolved (very
close to ZAMS) before contact, and therefore their mergers may be
replaced with ZAMS models with a mass of the parent binaries.
After coalescence, however, the mergers are much more massive and
their evolutions cannot be negected agian. To examine the effect
of AML in this cluster, we simply deal with low-mass binaries
($M_{\rm 1i} <1M_\odot$) as follows:

(1) From the binary sample, we found out some binaries with
$M_{\rm 1i}>1/2M_{\rm to}$ and with the orbital period less than
0.5 d at the cluster age by AML. The lower mass limit $1/2M_{\rm
to}$ is sure that the mergers are probably larger than the
turnoff, and the constrain on the orbital period leads the
binaries possibly be contact before or at the cluster age. A
semi-empirical formula for the orbital period variation is adopted
here \cite{ste06}:
\begin{equation}
$${\rm d} P_{\rm orb} \over {\rm d} t$$ = -(2.6+1.3) \times 10^{-10}P_{\rm orb}^{-1/3}e^{-0.2P_{\rm orb}}
\end{equation}
where $P_{\rm orb}$ in days and time in years. For very short
orbital periods the exponential factor is close to unity and
varies very little during the subsequent evolution of the orbital
period of the binaries. So it is ignored in this paper.

(2)Find out the time, $t_{\rm RLOF}$, at which the primary is full
of its Roche lobe for the selected binaries from step 1. A rapid
binary evolution code written by Hurley, Pols \& Tout (2002) is
employed here and the Roche lobe of the primary $R_{\rm cr1}$ is
calculated by \cite{egg83}
\begin{equation}
R_{\rm cr1}/A=$$ 0.49q^{2/3} \over 0.6q^{2/3}+{\rm
ln}(1+q^{1/3})$$ ,
\end{equation}
where A is the separation and $q=M_{\rm 1}/M_{\rm 2}$

(3)Evolve the mergers to the cluster age. The mergers are ZAMS
models instead with a mass of the systems. The starting age is
$t_{\rm RLOF}$, since the systems will {\bf reach} contact very
quickly from the onset of RLOF because of AML. Different
timescales $t_{\rm cc }$ are adopted here. Some of the binaries
from step 1 have left the main sequence, but most of them remain
on the main sequence.

Table 5 gives the results from different cluster age assumptions
and different $t_{\rm cc}$, and the magnitudes and colors for the
mergers from these cases are also presented in Fig. 11.

In Fig.\ref{montec}, we see that, the products from binary
evolution (AS and AR in the paper) and those from AML cover
different regions. The mergers from conservative evolution are
located in the region of high luminosity, i.e far away from the
the turnoff, and the remnants from AML occupy the region with low
luminosity and have a scatter on the color. The figure also
indicates that AML of low-mass binaries is much more important in
this old open cluster. In some cases (case 3 to 6), the produced
BSs from conservative evolutionary appears a certain width with
decreasing mass. Another phenomena worth noticing is in case 6,
the produced BSs result extends very close to F81, the observed
brightest BSs in this cluster. Further study shows that the lower
limit of the cluster age may be 3.8 Gyr for $t_{\rm
cc}=1\times10^9$ yr to result in the produced BSs extending to the
position of F81.

As shown in Table 4, we only obtained 38 BSs from a $10^6$ binary
sample in the widest condition. NGC 2682 is an open cluster and
has a much less stars than $10^6$. In Hurley et al.
\shortcite{hur05}, 12000 single stars and 12000 binaries
($N=24000$) are adopted in the best fitting model. Multipying this
factor, we just got 1 BS with high luminosity. There are 3 BSs
from the observations, however, located in the region covered by
the mergers from AS and AR. The number of stars initially in this
cluster is important here, but it is difficult for us to estimate
it because of the incompleteness of observations as determing the
current mass $M$ and the uncertainties of mass loss history of the
cluster when converting $M$ to an initial mass $M_{\rm 0}$. IMF
may also affect the results. We adopt the field single star IMF
for the primaries and the lower mass limit is $0.1M_\odot$ in the
binary sample, however, M67 is an old open cluster and rich in
binaries, and then it possibly has different IMF as well as
different lower mass limit as described by Chabrier
\shortcite{cha03}.

In the region with low luminosity, i.e the region covered by the
mergers of AML, the numbers shown in Table 5 are enough to account
for BSs in this region. Dynamically unstable mass transfer (AD) is
also possible to form BSs in this region, since the binaries
undergoing AD have larger initial mass ratios in general, leading
the smaller mergers. According to the result of Nelson \& Eggleton
(2001), we roughly estimate the number of binaries undergoing AD
evolution with initial primary mass between 0.89 and $1.26M_\odot$
as $N=10^6$. It is about 41--63 and the mass of the binaries are
less than  $1.36M_{\rm to}$, indicating that AD may produce BSs
near the turnoff indeed, but the contribution is {\bf much smaller
than that of AML in low-mass binaries.}

In  Hurley et al.\shortcite{hur05},
the authors obtained seven BSs from case A mass transfer leading to coalescence
by unperturbed evolution.
The age of the mergers in that paper were calculated based on the assumption
that core hydrogen burning proceeds uniformly and
that the end of the main sequence is reached when  10 per cent of the total
hydrogen has been burnt.
These assumptions for calculating the MS lives of the mergers
are questionable as they mentioned in the paper.
According to our calculations,
almost all of the mergers have relatively short main-sequence lives,
i. e. less than $10^9$ yr (see Fig.\ref{tms}).
It means that five of the seven BSs are likely to leave the main sequence
and no longer be observed as BSs at 4 Gyr.
The range of age for M67  is likely to improve the result.

\subsection{NGC 2660}
\begin{table*}
 \begin{minipage}{150mm}
 \caption{Consequences of Monte Carlo simulations from different assumptions
for binaries undergoing AS and AR evolutionary channel in NGC 2660.}
   \begin{tabular}{cccccc}
\hline
& conditions & $N_{\rm BS}$&$N_{\rm W UMa}$& $N_{\rm BS}$&$N_{\rm W UMa}$ \\
&&$(M_{\rm L}=0.8M_\odot)$&$(M_{\rm L}=0.8M_\odot)$&$(M_{\rm L}=0.1M_\odot)$&$(M_{\rm L}=0.1M_\odot)$\\
\hline
case 1 & $t_{\rm cluster}=1.2 \times 10^9$, $t_{\rm cc}=5 \times 10^7$  yr & 268 & 59 & 70 & 11 \\
\hline
case 2 & $t_{\rm cluster}=1.2 \times 10^9$, $t_{\rm cc}=1 \times 10^8$  yr &284&117&71 & 21\\
\hline
case 3 & $t_{\rm cluster}=1.2 \times 10^9$, $t_{\rm cc}=5 \times 10^8$  yr &296&631&72& 126\\
\hline
case 4 & $t_{\rm cluster}=1.2 \times 10^9$, $t_{\rm cc}=1 \times 10^9$  yr &87&1038&21&220\\
\hline
case 5 & $t_{\rm cluster}=0.9-1.2 \times 10^9$, $t_{\rm cc}=5 \times 10^7$  yr &693&454&141&87\\
\hline
case 6 & $t_{\rm cluster}=0.9-1.2 \times 10^9$, $t_{\rm cc}=1 \times 10^8$  yr &703&525&149&106\\
\hline
case 7 & $t_{\rm cluster}=0.9-1.2 \times 10^9$, $t_{\rm cc}=5 \times 10^8$  yr &445&934&102&194\\
\hline
case 8 & $t_{\rm cluster}=0.9-1.2 \times 10^9$, $t_{\rm cc}=1 \times 10^9$  yr &87&1125&21&241\\
\hline
\label{2660t}
\end{tabular}
\end{minipage}
\end{table*}

NGC 2660 is an intermediate-age open cluster with a solar
metallicity. The early study by Hartwick \& Hesser (1973) showed
the following properties for this cluster: $E(B-V)=0.38$,
$(m-M)_{\rm 0}=12.3 \pm 0.3$, age $\sim 1.2$ Gyr, metallicity
similar to the Hyades, and high possibility of membership for the
N-type carbon star. There are some uncertainties for the
determination of the cluster age, e.g. Lynga (1987) cited 1.6 Gyr,
Jane \& Phelps (1994) gave an age of 0.9 Gyr, while Carrio \&
Chiosi (1994) derived 0.7 Gyr. The latest report on this cluster
is Sandrelli et al. (1999). It was shown that, metallicity about
solar, $(m-M)_{\rm 0}=12.1 -12.3$, $E(B-V)=0.37-0.42$, age $ \le
1Gyr$, with a fraction of binaries of about 30 per cent.

According to the new catalogue of blue stragglers in open clusters
\cite{al07}, there are 8 BSs in this cluster while the cluster age
is ${\rm log}t=9.03$ (about 1.1 Gyr). The turnoff is around
$2M_\odot$ for this age, indicating that AML in low-mass binaries
has little influence on the product of BSs in this cluster. Table
6 presents our simulation results from different assumptions of
the cluster age and $t_{\rm cc}$ by interpolating in Table A1.

Though we show the simulation results for $t_{\rm cc}=5 \times
10^8$ and $t_{\rm cc}=1 \times 10^9$ yr in Table \ref{2660t}, we
should bear in mind that NGC 2660 is an intermediate-age cluster
and a long $t_{\rm cc}$ is unreasonable for it from both
observations and our models. Such a long  $t_{\rm cc}$ as $5
\times 10^8$ or $1 \times 10^9$ yr will inevitably cause large
differences of both components of a binary in our models, and
therefore a shorter $t_{\rm cc}$, namely, $t_{\rm cc}=1 \times
10^8$ and
 $t_{\rm cc}=5 \times 10^7$ yr,
is more appropriate for this cluster.
From Table \ref{2660t}
we see that BS number is not sensitive to $t_{\rm cc}$ in this cluster,
but the contact systems are very sensitive to $t_{\rm cc}$,
since a longer $t_{\rm cc}$
indicates a larger parameter space for these systems.
Contact binaries exist in the cluster for all the cases in Table \ref{2660t},
but it is less than the BSs number produced from AS and AR
when $t_{\rm cc}=1 \times 10^8$ and $t_{\rm cc}=5 \times 10^7$ yr,
especially in case 1,
it is just of 1/6 to 1/5 of that of BSs.

There are 150 stars on the main sequence to two magnitudes below the turnoff
in NGC 2660 and 175 in NGC 2682,
as shown in Ahumada \& Lapasset \shortcite{al07}.
From this comparison,
our models may produce the necessary BSs in
NGC 2660 when $M_{\rm L}=0.8M_\odot$.
If $M_{\rm L}=0.1M_\odot$ for the binary sample, however,
the produced BSs number from AS and AR is just 1/5 to 1/4 of that from
$M_{\rm L}=0.8M_\odot$.
Normalization is necessary here to estimate the BSs birthrate from our models
in this cluster
and the same problem mentioned in section 4.2 appears here again.
Nevertheless,
our models may explain several BSs in this intermediate age cluster.

\begin{figure*}
\centerline{\psfig{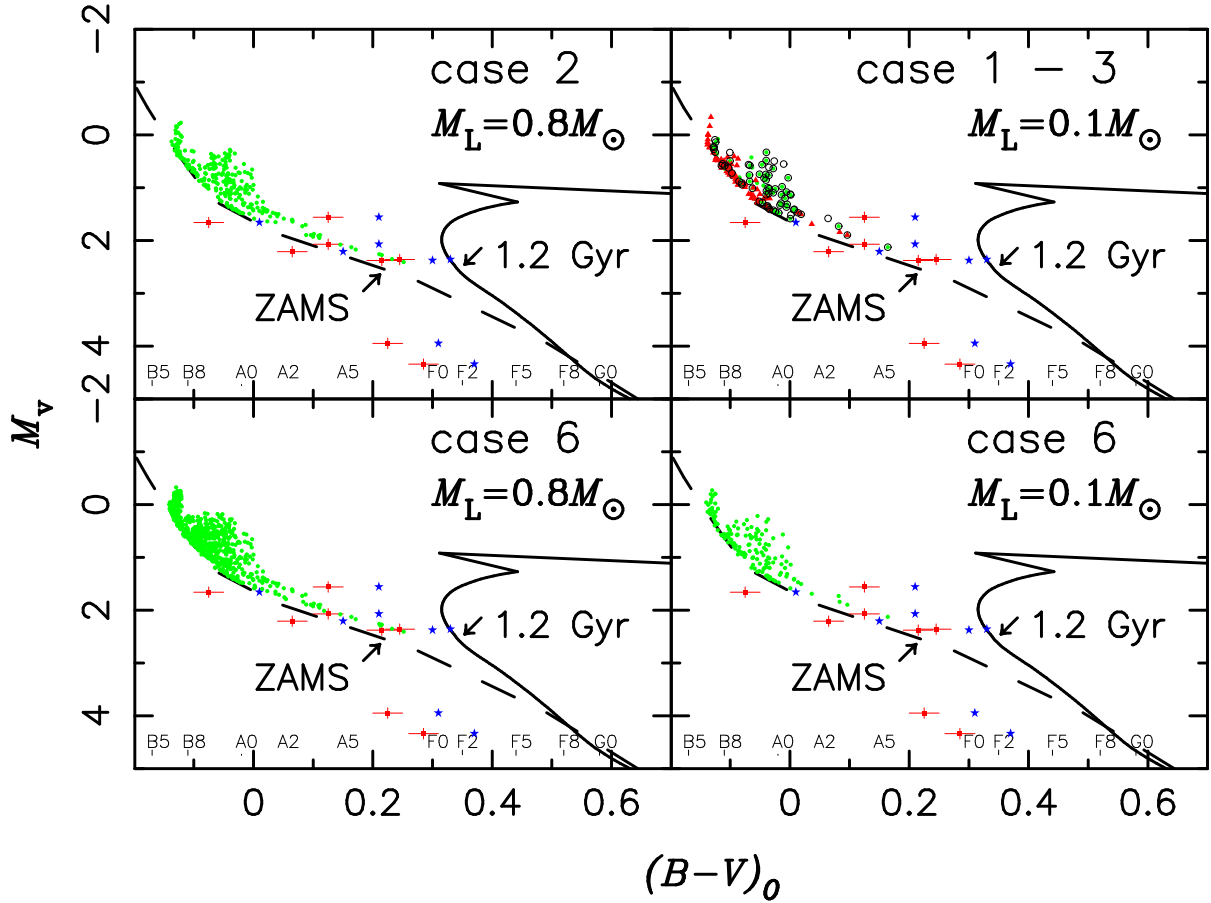}}
\caption{Monte-Carlo simulation results for NGC 2660 from
different assumptions (Table \ref{2660t}). The dots show the blue
stragglers listed by Ahumada \& Lapasset (2007), and the error
bars are from Sandrelli et a. (1999), i.e. $m-M=12.1-12.3$,
$E(B-V)=0.37-0.42$. The stars present the positions of BSs when
$m-M=12.2$, $E(B-V)=0.31$ as shown in Ahumada \& Lapasset (2007).
In the upper right panel, the circles, dots and triangles are for
$t_{\rm cc}$ equal to $5 \times 10^7$, $1 \times 10^8$ and $5
\times 10^8$, respectively. } \label{2660}
\end{figure*}

Figure \ref{2660} shows some examples from Monte Carlo simulation on CMD.
The observed BSs \cite{al07} and an isochrone ($t=1.2 Gyr$)
are also plotted in the figure.
Several BSs are located in the region of our models from AS and AR.
In the region with the most chance from our models $M_{\rm v} \approx 1.0$,
however,
there are no BSs observed at present.
The observed peak appears about $M_{\rm v} \approx 2.0$, one magnitude
lower than the theoretical value.
Mass loss during merger process may shorten the discrepancies,
but we should find out an appropriate mechanism to explain
such a large mass loss about $0.5M_\odot$,
which is much larger than that from observations
and from smooth particle hydrodynamic simulations.

Though the BSs number (or birthrate) from our models is not
sensitive to $t_{\rm cc}$, the initial parameter space will change
with  $t_{\rm cc}$, resulting in some differences for the mergers.
In the upper right panel, we show the results from case 1 to case
3 with $M_{\rm L}=0.1M_\odot$. As seen in this figure, the mergers
from a long $t_{\rm cc}$, i.e. $5\times 10^8$ yr, extend to a
higher luminosity and concentrated on the ZAMS, but small
differences appear between $t_{\rm cc}=1\times 10^8$ yr and
$t_{\rm cc}=5\times 10^7$ yr.

{\bf We have not considered the effect of AML in this cluster,
since we simply assume that the conservation assumptions are
reasonable for stars with spectra from G0 (about $1.05M_\odot$) to
B1 according to Nelson \& Eggleton \shortcite{nel01}. In fact, a
main sequence star can generate a magnetic field as long as it has
a sufficiently thick convective envelop, i.e. with a spectrum
later than F8. In the study of Andronov, Pinsonneault \& Terndrup
\shortcite{apt06}, the mass threshold for AML is between 1.2 and
$1.4M_\odot$, then the total mass of the merger might be up to 2.4
and $2.8M_\odot$, above the turnoff of this cluster. So it could
be possible that AML is also impacting the blue straggler
population in this intermediate-age cluster. This provides another
possible explanation for the magnitude offset from the models and
observations.}

\section{DISCUSSIONS AND CONCLUSIONS}
In Sect.4, we notice that the timescale from contact to complete
coalescence, $t_{\rm cc}$, strongly affects the initial parameter
space of primordial binaries which eventually produce single BSs
in a cluster. On the other hand, there are {\bf some conflicting
estimates} for $t_{\rm cc}$ from observations and theoretical
models. {\bf In this paper we adopt empirical values, i.e. $t_{\rm
cc}$ is short in comparison to the evolution timescale of both
components in a binary, and ignore the changes during the merger
process. In this section, we will first discuss the consequences
of a long $t_{\rm cc}$. Because of evolution of both components
during the merger process, the primaries have lower central
hydrogen content and the matter from the secondaries have larger
He content. The former results in a redder colour for the mergers
while the latter makes the mergers bluer. So the final positions
of the mergers are possibly similar to those shown in this paper,
except that the primaries have left the main sequence at final
coalescence. This case will appear in the mergers with little
central hydrogen. For example, a star with $2M_\odot$ may evolve
from ZAMS to exhausted of central hydrogen in $10^9$ yr, and then
none of the mergers from binaries with primary' masses larger than
$2M_\odot$ will be on the main sequence if $t_{\rm cc}= 1 \times
10^9$ yr. For the primaries with very little hydrogen in the
center at contact, the mergers may never be on the main sequence
even in the cases of short $t_{\rm cc}$. The long $t_{\rm cc}$
also delays the appearance of the mergers and shortens their
timescales on the main sequence. The latter has not been exactly
expressed in our models, and therefore we just see that the
mergers from a long $t_{\rm cc}$ have larger luminosities as shown
in section 4.}

{\bf In our binary evolutions, we have not included AML, which
exists in low-mass binaries and may be the main course making the
binaries change from detached to contact and finally coalesce,
resulting in a large contribution to BSs in old clusters, e.g NGC
2682. In young and intermediate-age clusters, however, AML has
little contribution to the birthrate of BSs, since (a) the time is
not long enough for binaries to go from detached to contact and
(b) the mass of the mergers is probably less than the turnoff of
the cluster even though their parents may coalesce in the cluster
age. So, we simply estimated the effect of AML in NGC 2682 while
negecting it in NGC 2660.}

The mass loss during the merger process can also affect our
result, mainly the location on the CMD of the products. As shown
in NGC 2660, no BSs have been observed in the region with the most
opportunity from our models. Because of mass loss, the mergers
will be fainter than those given in the paper. However the
faintness will be slight since the mass loss is not vast from both
observations and smooth particle hydrodynamic simulations
\cite{lrs96,sill97,sill01}. The lost mass may carry some angular
momentum out from the parent binary. By analyzing the BS spectra
from Hubble Space Telescope (there is an apparent continuum
deficit on the short-wavelength side of Balmer discontinuity ), De
Marco et al. \shortcite{dm04} argued that some BSs might be
surrounded by a circumstellar disk. However, Porter \& Townsend
\shortcite{pt05} showed that the flux deficits may be attributed
wholly to rapid rotation. The rotation rates needed are of the
order of those found in the study of De Macro et al.
\shortcite{dm05}. Whether the flux deficits shortward of the
Balmer jump are induced by a circumstellar disk or rapid rotation,
it provides a possible explanation for the orbital angular
momentum of the system after coalescence. Such a large mass loss
as shown in NGC 2660 (about $0.5 M_\odot$), however, is a problem
and should be explained reasonably in physics.

Based on some assumptions, we studied the mergers of close
binaries from AS and AR evolution by detailed evolutionary
calculations. The products from our models may stay on the left of
the ZAMS and have no central concentration with decreasing mass.
Because of the {\bf development} of the convective core, the
mergers with little central hydrogen (less than 0.01) in our
models have unusually long timescales on the main sequence ($\sim
10^8$ yrs). These objects are probably subgiants as they are
formed, since the primaries in the progenitors also have little
central hydrogen and may have left the main sequence during merger
process.

The mergers from our models stay on the main sequence for a
timescale in order of $10^8$ yrs. Some low-mass mergers may stay
on the MS for about $10^9$ yrs. The timescale is similar to that
of W UMa stars from observations, and therefore we may roughly
estimate the contribution to BSs from AS and AR via the number of
W UMa systems in a cluster. The estimation, however, is not
absolutely since both of the two timescales have wide ranges and
large uncertainties, and we cannot rule out other methods for
creating W UMa systems except for AS and AR. Comparison to
observations indicates that our models (binary coalescence from AS
and AR) are not important for the produce of BSs in old open
clusters, while likely play a critical role in some younger open
clusters.

We performed Monte Carlo simulations to examine our models in an
old open cluster NGC 2682 and in an intermediate-age cluster NGC
2660. The effect of AML was estimated in NGC 2682 in a simple way,
where the mergers are replaced with ZAMS models. In NGC 2682,
binary mergers from our models cover the region with high
luminosity and those from AML are located in the region with low
luminosity, existing a certain width. The BSs from AML are much
more than those from our models, indicating that AML of low mass
binaries makes a major contribution to BSs in this cluster. Our
models are corresponding for several BSs in NGC 2660. In the
region with the most opportunity on CMD, however, no BSs have been
observed. {\bf Our results are well-matched to the observations if
there are $\sim 0.5M_\odot$ of mass loss in the merger process,
but a physical mechanism for this much mass loss is a problem.}


\section{ACKNOWLEDGMENTS}
The authors thank R. S. Pokorny for his improvement in language.
This work is supported by
the Chinese National Science Foundation (Grant Nos. 06GJ061001 and 10433030),
the Yunnan Natural Science Foundation (Grant No. 2004A0022Q) and
the Chinese Academy of Sciences (Grant No. O6YQ011001).

\appendix
\section{Evolutionary results in the paper}
We choose a set of binaries undergoing AS and AR evolution
from Nelson \& Eggleton \shortcite{nel01}
to study the characteristics of the merger products and their
connections with blue stragglers.
Table A1 gives the initial parameters of the binary systems,
their RLOF information and
the structures and evolutionary consequences of the mergers.
The first three columns contain the initial mass of the primary
$M_{\rm 1i}$,
the initial mass ratio $q_{\rm i}$ (the primary to the secondary)
and the initial orbital period $P_{\rm i}$ in logarithmic, where
$P_{\rm ZAMS}$ is given by (Nelson \& Eggleton, 2001)
\begin{equation}
P_{\rm ZAMS} \approx $$0.19M_{\rm 1i}+0.47M_{\rm 1i}^{2.33}\over 1+1.18M_{\rm 1i}^2$$.
\end{equation}
The fourth and fifth columns are the ages at which
Roche lobe overflow begins ($t_{\rm RLOF}$)
and the binary comes into contact ($t_{\rm contact}$) in our calculation.
The next three columns show some system parameters at $t_{\rm contact}$,
i.e. the mass of the primary $M_{\rm 1}$,
the mass of the secondary $M_{\rm 2}$
and the orbital period $P_{\rm contact}$.
The remaining columns present the evolutionary results of the mergers:
the lifetime on the main sequence ($t_{\rm MS}$),
the central hydrogen mass fraction of the merger as constructed
($X_{\rm Hcc}$) and after adjustment ($X_{\rm Hcm}$),
surface abundances for the elements H ($X_{\rm Hs}$), He ($X_{\rm Hes}$)
and the ratio of C/N at the surface ($(C/N)_{\rm s}$).

\end{document}